\input harvmac
\input epsf

\input amssym.tex

\def\a{\alpha}
\def\b{\beta}
\def\c{\gamma}
\def\C{\Gamma}

\def\D{\Delta}

\def\vt{\vartheta}
\def\k{\kappa}
\def\l{\lambda}
\def\L{\Lambda}
\def\m{\mu}
\def\n{\nu}
\def\r{\rho}
\def\s{\sigma}

\def\t{\tau}
\def\th{\theta}

\def\sq{\sqrt{1-4\m^2}}

\def\cJ{{\cal J}}

\def\cE{{\cal E}}

\def\cH{{\cal H}}

\def\cO{{\cal O}}

\def\cN{{\cal N}}
\def\cQ{{\cal Q}}

\def\cP{{\cal P}}
\def\cV{{\cal V}}

\def\ie{i.e.~}
\def\eg{e.g.~}
\def\cf{cf.~}

\Title{\vbox{\hbox{hep-th/0402092}\hbox{UU-04-04}}}{\vbox{\centerline{Higher Conserved Charges and Integrability}
\medskip \centerline{for Spinning Strings in AdS$_5\times$S$^5$}}}

\centerline{{J. Engquist}\footnote{$^\dagger$}
{\tt johan.engquist@teorfys.uu.se}}
\bigskip\centerline{\it Department of Theoretical Physics}
\centerline{\it Uppsala University}\centerline{\it Uppsala, Box 803, SE-751 08, Sweden}

\vskip .3in

\centerline{\bf Abstract}

\vskip .1in

We demonstrate the existence of an infinite number of local commuting charges for classical solutions of the string sigma 
model on AdS$_5\times$S$^5$ associated with a certain circular three-spin solution spinning with large angular momenta in
three orthogonal directions on the five-sphere. Using the AdS/CFT correspondence we find agreement to one-loop with
the tower of conserved higher charges in planar $\cN=4$ super Yang-Mills theory associated with the dual composite
single-trace operator in the highest weight representation $(J_1,J_2,J_2)$ of $SO(6)$. The agreement can be explained by the
presence of integrability on both sides of the duality.

\Date{February 2004} 


\lref\AldayZB{
L.~F.~Alday,
{\it ``Nonlocal Charges on AdS$_5$  $\times$ S$^5$ and pp-waves},''
JHEP {\bf 0312}, 033 (2003), {\tt hep-th/0310146}.
}

\lref\ArutyunovRG{
G.~Arutyunov and M.~Staudacher,
{\it ``Matching Higher Conserved Charges for Strings and Spins,}'', {\tt hep-th/0310182}.
}

\lref\ArutyunovUJ{
G.~Arutyunov, S.~Frolov, J.~Russo and A.~A.~Tseytlin,
{\it``Spinning Strings in AdS$_5$  $\times$ S$^5$ and Integrable Systems,}'',
Nucl.\ Phys.\ B {\bf 671}, 3 (2003) {\tt hep-th/0307191}.
}

\lref\ArutyunovZA{
G.~Arutyunov, J.~Russo and A.~A.~Tseytlin,
{\it``Spinning Strings in AdS$_5$  $\times$ S$^5$: New Integrable System Relations,}''
{\tt hep-th/0311004}.
}

\lref\BeisertEA{
N.~Beisert, S.~Frolov, M.~Staudacher and A.~A.~Tseytlin,
{\it``Precision Spectroscopy of AdS/CFT,}''
JHEP {\bf 0310}, 037 (2003) , {\tt hep-th/0308117}.
}

\lref\BeisertYB{
N.~Beisert and M.~Staudacher,
{\it``The $\cN$ = 4 SYM Integrable Super Spin Chain,}''
Nucl.\ Phys.\ B {\bf 670}, 439 (2003), {\tt hep-th/0307042}.
}

\lref\BeisertXU{
N.~Beisert, J.~A.~Minahan, M.~Staudacher and K.~Zarembo,
{\it``Stringing Spins and Spinning Strings,}''
JHEP {\bf 0309}, 010 (2003), {\tt hep-th/0306139}.
}

\lref\BeisertTQ{
N.~Beisert, C.~Kristjansen and M.~Staudacher,
{\it``The Dilatation Operator of $\cN$ = 4 Super Yang-Mills Theory,}''
Nucl.\ Phys.\ B {\bf 664}, 131 (2003), {\tt hep-th/0303060}.
}

\lref\BeisertFF{
N.~Beisert, C.~Kristjansen, J.~Plefka and M.~Staudacher,
{\it``BMN Gauge Theory as a Quantum Mechanical System,}''
Phys.\ Lett.\ B {\bf 558}, 229 (2003), {\tt hep-th/0212269}.
}

\lref\BeisertBB{
N.~Beisert, C.~Kristjansen, J.~Plefka, G.~W.~Semenoff and M.~Staudacher,
{\it``BMN Correlators and Operator Mixing in $\cN$ = 4 Super Yang-Mills theory,}''
Nucl.\ Phys.\ B {\bf 650}, 125 (2003), {\tt hep-th/0208178}.
}

\lref\BeisertYS{
N.~Beisert,
{\it``The su(2$|$3) Dynamic Spin Chain,}'' {\tt hep-th/0310252}.
}

\lref\BeisertJB{
N.~Beisert,
{\it``Higher Loops, Integrability and the Near BMN Limit,}''
JHEP {\bf 0309}, 062 (2003), {\tt hep-th/0308074}.
}

\lref\BeisertJJ{
N.~Beisert,
{\it ``The Complete One-loop Dilatation Operator of $\cN$ = 4 Super Yang-Mills
theory,}''
Nucl.\ Phys.\ B {\bf 676}, 3 (2004), {\tt hep-th/0307015}.
}

\lref\BelitskyQH{
A.~V.~Belitsky,
{\it ``Fine Structure of Spectrum of Twist-three Operators in {QCD},}''
Phys.\ Lett.\ B {\bf 453}, 59 (1999), {\tt hep-ph/9902361}.
}

\lref\BelitskyRU{
A.~V.~Belitsky,
{\it ``Integrability and WKB Solution of Twist-three Evolution Equations,}''
Nucl.\ Phys.\ B {\bf 558}, 259 (1999), {\tt hep-ph/9903512}.
}

\lref\BelitskyBF{
A.~V.~Belitsky,
{\it ``Renormalization of Twist-three Operators and Integrable Lattice Models,}''
Nucl.\ Phys.\ B {\bf 574}, 407 (2000), {\tt hep-ph/9907420}.
}

\lref\BelitskyYS{
A.~V.~Belitsky, A.~S.~Gorsky and G.~P.~Korchemsky,
{\it``Gauge / String Duality for QCD Conformal Operators,}''
Nucl.\ Phys.\ B {\bf 667}, 3 (2003), {\tt hep-th/0304028}.
}

\lref\BenaWD{
I.~Bena, J.~Polchinski and R.~Roiban,
{\it``Hidden Symmetries of the AdS$_5\times$ S$^5$ Superstring,}'' {\tt hep-th/0305116}.
}

\lref\BerensteinJQ{
D.~Berenstein, J.~M.~Maldacena and H.~Nastase,
{\it``Strings in Flat Space and pp waves from $\cN$ = 4 Super Yang Mills,}''
JHEP {\bf 0204}, 013 (2002), {\tt hep-th/0202021}.
}

\lref\BernardYA{
D.~Bernard,
{\it ``An Introduction to Yangian Symmetries,}''
Int.\ J.\ Mod.\ Phys.\ B {\bf 7}, 3517 (1993), {\tt hep-th/9211133}.
}

\lref\BraunID{
V.~M.~Braun, S.~E.~Derkachov and A.~N.~Manashov,
{\it ``Integrability of Three-particle Evolution Equations in {QCD},}''
Phys.\ Rev.\ Lett.\  {\bf 81}, 2020 (1998), {\tt hep-ph/9805225}.
}

\lref\BraunTE{
V.~M.~Braun, S.~E.~Derkachov, G.~P.~Korchemsky and A.~N.~Manashov,
{\it ``Baryon Distribution Amplitudes in {QCD},}''
Nucl.\ Phys.\ B {\bf 553}, 355 (1999), {\tt hep-ph/9902375}.
}

\lref\CallanXR{
C.~G.~.~Callan, H.~K.~Lee, T.~McLoughlin, J.~H.~Schwarz, I.~Swanson and X.~Wu,
{\it``Quantizing String Theory in AdS$_5\times$ S$^5$: Beyond the pp-wave,}''
Nucl.\ Phys.\ B {\bf 673}, 3 (2003), {\tt hep-th/0307032}.
}

\lref\DerkachovZE{
S.~E.~Derkachov, G.~P.~Korchemsky and A.~N.~Manashov,
{\it ``Evolution Equations for Quark Gluon Distributions in Multi-color QCD and
Open Spin Chains,}''
Nucl.\ Phys.\ B {\bf 566}, 203 (2000), {\tt hep-ph/9909539}.
}

\lref\DolanUH{
L.~Dolan, C.~R.~Nappi and E.~Witten,
{\it``A Relation Between Approaches to Integrability in Superconformal Yang-Mills
theory,}''
JHEP {\bf 0310}, 017 (2003), {\tt hep-th/0308089}.
}

\lref\DolanPS{
L.~Dolan, C.~R.~Nappi and E.~Witten,
{\it ``Yangian Symmetry in $D=4$ Superconformal Yang-Mills Theory,}'' {\tt hep-th/0401243}.
}

\lref\EngquistRN{
J.~Engquist, J.~A.~Minahan and K.~Zarembo,
``{\it Yang-Mills Duals for Semiclassical Strings on AdS$_5\times$ S$^5$,}''
JHEP {\bf 0311}, 063 (2003), {\tt hep-th/0310188}.}

\lref\EynardCN{
B.~Eynard and J.~Zinn-Justin,
{\it``The O(n) Model on a Random Surface: Critical Points and Large Order
Behavior,}''
Nucl.\ Phys.\ B {\bf 386}, 558 (1992),  {\tt hep-th/9204082}.
}

\lref\FaddeevZG{
L.~D.~Faddeev and G.~P.~Korchemsky,
{\it ``High-energy QCD as a Completely Integrable Model,}''
Phys.\ Lett.\ B {\bf 342}, 311 (1995), {\tt hep-th/9404173}.
}

\lref\FaddeevIY{
L.~D.~Faddeev,
{\it``How Algebraic Bethe Ansatz works for Integrable Model,}'' {\tt hep-th/9605187}.
}

\lref\FrolovAV{
S.~Frolov and A.~A.~Tseytlin,
{\it``Semiclassical Quantization of Rotating Superstring in AdS$_5$  $\times$ S$^5$,}''
JHEP {\bf 0206}, 007 (2002), {\tt hep-th/0204226}.
}

\lref\FrolovQC{
S.~Frolov and A.~A.~Tseytlin,
{\it``Multi-spin String Solutions in AdS$_5$  $\times$ S$^5$,}''
Nucl.\ Phys.\ B {\bf 668}, 77 (2003), {\tt hep-th/0304255}.
}

\lref\FrolovTU{
S.~Frolov and A.~A.~Tseytlin,
{\it ``Quantizing Three-spin String Solution in AdS$_5$  $\times$ S$^5$,}''
JHEP {\bf 0307}, 016 (2003), {\tt hep-th/0306130}.
}

\lref\FrolovXY{
S.~Frolov and A.~A.~Tseytlin,
{\it ``Rotating String Solutions: AdS/CFT Duality in Non-supersymmetric Sectors,}''
Phys.\ Lett.\ B {\bf 570}, 96 (2003), {\tt hep-th/0306143}.
}

\lref\GorskyNQ{
A.~Gorsky,
{\it ``Spin Chains and Gauge / String Duality,}'' {\tt hep-th/0308182}.
}

\lref\GubserTV{
S.~S.~Gubser, I.~R.~Klebanov and A.~M.~Polyakov,
{\it ``A Semi-classical Limit of the Gauge/String Correspondence,}''
Nucl.\ Phys.\ B {\bf 636}, 99 (2002), {\tt hep-th/0204051}.
}

\lref\KhanSM{
A.~Khan and A.~L.~Larsen,
{\it``Spinning Pulsating String Solitons in AdS$_5$  $\times$ S$^5$,}''
Phys.\ Rev.\ D {\bf 69}, 026001 (2004), {\tt hep-th/0310019}.
}

\lref\KostovPN{
I.~K.~Kostov and M.~Staudacher,
{\it ``Multicritical Phases Of The $O(N)$ Model On A Random Lattice,}''
Nucl.\ Phys.\ B {\bf 384}, 459 (1992), {\tt hep-th/9203030}.
}

\lref\KristjansenEI{
C.~Kristjansen,
{\it ``Three-spin Strings on AdS$_5$  $\times$ S$^5$ from $\cN=4$ SYM,}'' {\tt hep-th/0402033}.
}

\lref\KruczenskiGT{
M.~Kruczenski,
{\it ``Spin Chains and String Theory,}'' {\tt hep-th/0311203}.
}

\lref\LipatovYB{
L.~N.~Lipatov,
{\it ``High-energy Asymptotics of Multicolor QCD and Exactly Solvable Lattice
models,}''
JETP Lett.\  {\bf 59}, 596 (1994)
[Pisma Zh.\ Eksp.\ Teor.\ Fiz.\  {\bf 59}, 571 (1994)], {\tt hep-th/9311037}.
}

\lref\MandalFS{
G.~Mandal, N.~V.~Suryanarayana and S.~R.~Wadia,
{\it ``Aspects of Semiclassical Strings in AdS$_5$,}''
Phys.\ Lett.\ B {\bf 543}, 81 (2002), {\tt hep-th/0206103}.
}

\lref\MateosDE{
D.~Mateos, T.~Mateos and P.~K.~Townsend,
{\it ``Supersymmetry of Tensionless Rotating Strings in AdS$_5$  $\times$ S$^5$, and
Nearly-BPS Operators,}''
JHEP {\bf 0312}, 017 (2003), {\tt hep-th/0309114}.
}

\lref\MateosRN{
D.~Mateos, T.~Mateos and P.~K.~Townsend,
{\it ``More on Supersymmetric Tensionless Rotating Strings in AdS$_5$  $\times$ S$^5$,}''
{\tt hep-th/0401058}.
}

\lref\MikhailovGQ{
A.~Mikhailov,
{\it ``Speeding Strings,}''
JHEP {\bf 0312}, 058 (2003), {\tt hep-th/0311019}.
}

\lref\MikhailovQF{
A.~Mikhailov,
{\it``Slow Evolution of Nearly-degenerate Extremal Surfaces,}'' ~{\tt hep-th/0402067}.
}

\lref\MinahanRC{
J.~A.~Minahan,
{\it ``Circular Semiclassical String Solutions on AdS$_5$  $\times$ S$^5$,}''
Nucl.\ Phys.\ B {\bf 648}, 203 (2003), {\tt hep-th/0209047}.
}

\lref\MinahanVE{
J.~A.~Minahan and K.~Zarembo,
{\it ``The Bethe-ansatz for $\cN$ = 4 Super Yang-Mills,}''
JHEP {\bf 0303}, 013 (2003), {\tt hep-th/0212208}.
}

\lref\OgielskiHV{
A.~T.~Ogielski, M.~K.~Prasad, A.~Sinha and L.~L.~Wang,
{\it``B\"acklund Transformations and Local Conservation Laws for Principal Chiral
Fields,}''
Phys.\ Lett.\ B {\bf 91}, 387 (1980).
}

\lref\PohlmeyerNB{
K.~Pohlmeyer,
{\it ``Integrable Hamiltonian Systems and Interactions Through Quadratic
Constraints,}''
Commun.\ Math.\ Phys.\  {\bf 46}, 207 (1976).
}

\lref\ReshetikhinVW{
N.~y.~Reshetikhin,
{\it ``A Method of Functional Equations in the Theory of Exactly Solvable Quantum
Systems,}''
Lett.\ Math.\ Phys.\  {\bf 7}, 205 (1983).
}

\lref\ReshetikhinVD{
N.~Y.~Reshetikhin,
{\it ``Integrable Models of Quantum One-Dimensional Magnets with O(N) and Sp(2k)
Symmetry,}''
Theor.\ Math.\ Phys.\  {\bf 63}, 555 (1985)
[Teor.\ Mat.\ Fiz.\  {\bf 63}, 347 (1985)].
}

\lref\RussoSR{
J.~G.~Russo,
{\it ``Anomalous Dimensions in Gauge Theories from Rotating Strings in AdS$_5$ $\times$ S$^5$,}''
JHEP {\bf 0206}, 038 (2002), {\tt hep-th/0205244}.
}

\lref\SerbanJF{
D.~Serban and M.~Staudacher,
{\it ``Planar $\cN$ = 4 Gauge Theory and the Inozemtsev Long Range Spin Chain,}''
{\tt hep-th/0401057}.
}

\lref\TseytlinII{
A.~A.~Tseytlin,
{\it``Spinning Strings and AdS/CFT Duality,}''
{\tt hep-th/0311139}.
}

\lref\TseytlinNY{
A.~A.~Tseytlin,
{\it ``Semiclassical Quantization of Superstrings: AdS$_5$ $\times$ S$^5$ and Beyond},''
Int.\ J.\ Mod.\ Phys.\ A {\bf 18}, 981 (2003), {\tt hep-th/0209116}.
}

\lref\ValliloNX{
B.~C.~Vallilo,
{\it ``Flat Currents in the Classical AdS$_5$ $\times$ S$^5$ Pure Spinor Superstring,}''
{\tt hep-th/0307018}.
}

\lref\deVegaYZ{
H.~J.~de Vega, A.~L.~Larsen and N.~Sanchez,
{\it ``Semiclassical Quantization of Circular Strings in de Sitter and Anti-de
Sitter Space-times,}''
Phys.\ Rev.\ D {\bf 51}, 6917 (1995), {\tt hep-th/9410219}.
}

\newsec{Introduction}
The recent realization that the planar one-loop dilatation operator acting on composite gauge invariant operators in
$\cN=4$ $SU(N)$ super Yang-Mills theory (SYM) can be mapped to the Hamiltonian of an $SO(6)$ spin chain \MinahanVE, and
its supersymmetric extension \BeisertYB, has uncovered new structures hidden in certain sectors of the gauge theory, most
notably integrability. The AdS/CFT correspondence leads one to believe that a similar structure exists on the string theory
side, since the scaling dimensions of the operators in the gauge theory are identified with the energies of dual string 
states. Indeed, in \refs{\ArutyunovUJ,\ArutyunovZA} (see also \refs{\GorskyNQ,\MikhailovGQ,\KruczenskiGT,\MikhailovQF})\ it was
found that the $SO(4,2) \times SO(6)$ sigma model (which is known to be classically integrable) evaluated on a particular type of
rotating semiclassical string solutions collapses to a one-dimensional Neumann-Rosochatius (NR) system which is known to be
integrable. The integrability appears in a rather different way than on the gauge theory side, see \TseytlinII\ for a review.
Other evidence of integrability in type IIB string theory on AdS$_5\times$S$^5$ was found in 
\refs{\BenaWD,\DolanUH,\MandalFS,\AldayZB,\ValliloNX,\DolanPS}. Furthermore, integrable structures in gauge theories have been
found previously in \eg QCD \refs{\LipatovYB\FaddeevZG\BraunID\BelitskyQH\BraunTE\BelitskyRU\BelitskyBF\DerkachovZE{--}\BelitskyYS},
but their relation to string theory is not clear.

The energies of the semiclassical rotating string states in the bulk carrying large $SO(6)$ angular momenta
$J_I$, $I=1,2,3$, admit an expansion in the effective coupling constant $\l/J^2=R^4/\a'^2J^2\ll1$, where 
$J=J_1+J_2+J_3$, and have been argued to receive no $\a'$ corrections in the strict $J\rightarrow\infty$ limit, with 
$\l/J^2$ kept fixed, 
\refs{\BerensteinJQ,\GubserTV,\FrolovAV}. Furthermore, these states have been conjectured \FrolovQC\ to be dual in
the planar limit to holomorphic single-trace operators of the form
\eqn\operator{\cO={\rm tr}(\Phi_1+i\Phi_2)^{J_1}(\Phi_3+i\Phi_4)^{J_2}(\Phi_5+i\Phi_6)^{J_3}+\cdots\ , }
where $\Phi_A$, $A=1,\ldots,6$ are the real adjoint scalars in $\cN=4$ $SU(N)$ SYM. The holomorphicity requirement on the
operators can be relaxed \EngquistRN\ if \eg pulsating string solutions are considered 
\refs{\EngquistRN,\deVegaYZ,\GubserTV,\MinahanRC,\KhanSM}.
The dots in \operator\ indicate that to one loop all operators of the same canonical dimension built of only scalars mix among
themselves under renormalization, and need to be diagonalized in order to acquire well-defined scaling dimensions. In
general this is a highly non-trivial problem when considering operators for which the number of constituent fields $J$ in
\operator\ approaches infinity. However, due to the mapping of the one-loop dilatation operator to the Hamiltonian of a
spin chain the diagonalization procedure can be accomplished with Bethe ansatz techniques \FaddeevIY\ which 
often simplifies in this ``thermodynamic limit'', when the number of lattice sites on the spin chain approches infinity. 

Scaling dimensions of operators of the form given in \operator\ as well as of non-holomorphic operators with
insertions of anti-holomorphic ``impurities'' have been found in several examples using the Bethe ansatz -- to one loop in 
\refs{\MinahanVE,\BeisertXU,\BeisertEA,\EngquistRN,\KristjansenEI}\ and to two loops in \SerbanJF. These have been
successfully matched to the energies of the dual string states calculated in \refs{\BerensteinJQ,\GubserTV,\FrolovAV,
\FrolovQC,\EngquistRN,\MinahanRC,\TseytlinNY,\RussoSR,\FrolovTU,\FrolovXY}. There are some indications, however, that this
impressive agreement starts to break down at three loops \refs{\CallanXR,\SerbanJF}. 

The integrability on the gauge theory side implies as many integrals of motion as there are constituent elementary
fields in the (scalar) operator under consideration, which is equal to the total $R$-charge $J$ in the case of holomorphic
operators. In particular, in the thermodynamic limit an infinite number of conserved charges is expected.
In \EngquistRN\ and \ArutyunovRG\ it was independently realized that the generator of charges in the thermodynamic limit can
be identified with the resolvent of the Bethe roots. Focusing on the 
S$^5$ part, by applying known B\"acklund transformations for $O(n)$ sigma models \refs{\PohlmeyerNB,\OgielskiHV},
these charges were shown to be present also on the string theory side \ArutyunovRG\ for a certain type of folded 
and circular two-spin string solutions \ArutyunovUJ\ by constructing the generator of conserved higher charges in the
semiclassical limit and extracting the $\cO(\l)$ contribution. The aim of the present work is to determine the generator of
conserved higher charges associated with a circular three-spin solution \refs{\FrolovQC,\FrolovTU}\ for which two of 
the $R$-charges are equal, say $J_2=J_3$. This state belongs to a subset of the space of states of the NR system
\ArutyunovZA\ having a constant Lagrange multiplier $\L$. For these states one can show that the relation
$\sum_I m_I J_I=0$ must hold for some integers $m_I$, $I=1,2,3$, which for the $(J_1,J_2,J_2)$ state can be chosen to 
be $m_1=0$, $m_2=-m_3=m$. We will find that the generator satisfies a sixth order equation. By restricting to 
one loop, this sixth order equation factorizes into two cubic equations, one of which is identical to
the cubic equation satisfied by the generator on the gauge theory side found in a previous work \EngquistRN! The 
other cubic equation is related to the first one by a discrete transformation.

The paper is organized as follows. In the next section we present a review of the gauge theory calculations,
valid to one loop, leading to the infinite set of local commuting charges associated with the operator \operator\ in
the $SO(6)$ representation $[0,J_1-J_2,2J_2]$, $J_1>J_2$. In Section 3 we summarize the classical three-spin string
solution dual to this operator and determine the associated generator of commuting charges by solving a set of B\"acklund
equations first considered in the context of type IIB string theory on $S^5$ in \ArutyunovRG. Some of the details of
these calculations are collected in Appendices A and B. Finally, in Section 4 we give our conclusions. 

\newsec{Gauge Theory Calculations}
In this section we review the planar $\cN=4$ $SU(N)$ SYM calculations presented in \EngquistRN\ leading to the generator of
conserved charges associated with the operator given in \operator\ having large $R$-charges, valid to one loop in 
$\l=g_{\rm YM}^2N$.
There, it was argued that the set of Bethe equations for which $J_1>J_2=J_3$ and $J_1<J_2=J_3$ need to be considered 
separately, and reduce respectively to certain types of integral equations found in $O(-1)$ and $O(+1)$ matrix models, see \eg
\refs{\KostovPN,\EynardCN}. Furthermore, it was asserted that the generator in the sector with $J_1>J_2=J_3$ coincides with the
one in the sector $J_1<J_2=J_3$, so for simplicity we shall focus on the former case. 

\subsec{The One-loop Matrix of Anomalous Dimensions and the Bethe Ansatz}
We start by presenting the general structure of the planar {\it one-loop} mixing matrix acting on single-trace composite 
operators constructed out of $J$ of the three adjoint complex scalars of the $SU(N)$ $\cN=4$ SYM supermultiplet. 
Restricting to $SO(6)$, the one-loop matrix of anomalous dimensions can be written in the operator form \MinahanVE
\eqn\mixing{\C={\l \over 16\pi^2} \sum_{\ell=1}^{J}\big(K_{\ell,\ell+1}+2{\rm 1\kern-.33em 1}_{\ell,\ell+1}
-2P_{\ell,\ell+1}\big)\ , }
acting on a $6^J$-dimensional Hilbert space $\cH=\bigotimes_{\ell=1}^{J}\cH_\ell$, each module carrying the vector 
representation of $SO(6)$. $K_{\ell,\ell+1}$ and $P_{\ell,\ell+1}$ are trace and permutation operators respectively 
acting non-trivially only on the modules $\cH_{\ell}$ and $\cH_{\ell+1}$. As usual, a given set of operators having the same 
bare dimensions will generically mix under renormalization and needs to be diagonalized to produce eigenstates of the 
dilatation operator. We recall that for one-loop renormalization there is no mixing of operators of the form \operator\ 
containing only scalars with operators containing gluons or fermions. For our purposes, as we analyze operators with large 
$R$-charges, the diagonalization procedure can be considerably simplified by recognizing \MinahanVE\ equation
\mixing\ as the Hamiltonian of an $SO(6)$ closed quantum spin chain with nearest-neighbor interactions between the $J$ lattice
sites\foot{The {\it complete} one-loop
dilatation operator of $\cN=4$ SYM \BeisertJJ\ is mapped to the Hamiltonian of an $PSU(2,2|4) \supset SO(6)\times SO(4,2)$
super spin chain \BeisertYB. }. The spin chain is characterized by the Bethe eigenstates which in the gauge theory correspond
to diagonal composite operators possessing well-defined scaling
dimensions. Moreover, the Bethe eigenstates are completely specified by their distributions of Bethe roots that solve the
following set of Bethe equations \refs{\ReshetikhinVW,\ReshetikhinVD}:
\eqn\betheeqs{\eqalign{\Big({u_{1,i}+i/2 \over u_{1,i}-i/2}\Big)^J&=
\prod_{j\ne i}^{n_1}{u_{1,i}-u_{1,j}+i \over u_{1,i}-u_{1,j}-i}
\prod_{j}^{n_2}{u_{1,i}-u_{2,j}-i/2 \over u_{1,i}-u_{2,j}+i/2} 
\prod_{j}^{n_3}{u_{1,i}-u_{3,j}-i/2 \over u_{1,i}-u_{3,j}+i/2} \ , \cr 
1&=\prod_{j\ne i}^{n_{2}}{u_{2,i}-u_{2,j}+i \over u_{2,i}-u_{2,j}-i}
\prod_{j}^{n_{1}}{u_{2,i}-u_{1,j}-i/2 \over u_{2,i}-u_{1,j}+i/2} \ , \cr 
1&=\prod_{j\ne i}^{n_{3}}{u_{3,i}-u_{3,j}+i \over u_{3,i}-u_{3,j}-i}
\prod_{j}^{n_{1}}{u_{3,i}-u_{1,j}-i/2 \over u_{3,i}-u_{1,j}+i/2}\ , } }
where three types of Bethe roots $u_I$, $I=1,2,3$, are needed, each associated with a simple root of the $SO(6)$ Lie algebra. 
In \betheeqs, $n_I$, $I=1,2,3$ denote the numbers of each type of root for a particular solution and are related to the angular 
momenta of the composite operator. In particular, for a generic solution carrying three independent angular 
momenta the relation is $(J_1,J_2,J_3)=(J-n_1,n_1-n_2-n_3,n_2-n_3)$. Note that $n_3=0$ for holomorphic operators which saturate
the following bound between the canonical dimension $\D_0$ and the $R$-charge (\cf \refs{\MateosDE,\MateosRN}): 
\eqn\bound{\D_0\geq J=J_1+J_2+J_3\ .}

The $SO(6)$ spin chain is known to be integrable, and we therefore expect the Hamiltonian in \mixing\ to be only one
particular linear combination of a set of $J$ commuting integrals of motion. After diagonalization the eigenvalue of the
generator of these commuting charges becomes
\FaddeevIY\ \eqn\gendef{t(u)= i\log\Big(\prod_k^{n_1}{u-u_{1,k}+i/2 \over u-u_{1,k}-i/2}\Big)\ , }
and is accordingly a polynomial in a spectral parameter $u$: \eqn\expand{t(u)=\sum_mt_mu^m. } The
coefficients $t_m$ are interpreted as higher commuting {\it local} charges if $t(u)$ is expanded around $u=0$. In particular,
the lowest charges $t_0$ and $t_1$ correspond respectively to the total momentum $P=\sum_kp_k$ and the energy eigenvalue $E$ of 
a quantum eigenstate of the spin chain. Due to the cyclicity of the trace we require $P=0$.  We can now fix the constant
of proportionality between the one-loop anomalous dimension and the energy of the spin chain to be
\eqn\hamilton{\c={\l \over 8\pi^2}t_1\ , } 
which expressed in terms of the $u_1$ Bethe roots becomes 
\eqn\andimm{\c={\l \over 8\pi^2}\sum_k^{n_1}{1 \over (u_{1,k})^2+1/4}\ , } obtained by expanding \gendef\ to first order in $u$.

\subsec{The Gauge Dual of the Frolov-Tseytlin String}
For the holomorphic operator we are analyzing, \ie the operator \operator\ with $J_1>J_2=J_3$ and with $R$-charge assignments
$(J_1,J_2,J_2)$, or $[0,J_1-J_2,2J_2]$ in terms of Dynkin labels, only two types of Bethe roots are needed, as previously 
mentioned. This highest weight representation actually corresponds to the case of ``half-filling'' where the number of 
$u_2$ roots equals half the number of $u_1$ roots \EngquistRN. In terms of the filling 
fractions $\a\equiv n_1/J$ and $\b\equiv n_2/J$ this amounts to the condition $\a=2\b$. By taking the thermodynamic limit, 
\ie by letting $J=\sum_IJ_I\rightarrow \infty$, the Bethe roots accumulate on smooth contours and it is convenient to
introduce distribution densities of the Bethe roots, which we normalize as:
\eqn\unity{\int_{\CC_+}dv\s(v)=\int_{\CC_2}dv\r_2(v)=1\ .} The $u_1$ roots
are assumed to be distributed symmetrically around $u=0$ on two disjoint contours $\CC_{\pm}$ intersecting the real axis,
where $\CC_-$ is the mirror
image of $\CC_+$. Furthermore, the $u_2$ roots are assumed to accumulate on a single curve $\CC_2$ on the imaginary axis
symmetrically around $u=0$. Let us rewrite the $SO(6)$ Bethe 
equations in the thermodynamic limit. By rescaling $u$ by $J$ and taking logarithms, \betheeqs\ reduces to a set of singular 
integral equations  
\eqn\singeqone{ {1 \over u}-2\pi m= \alpha{\int_{\CC_+}\kern-1.63em -}~~dv{\s(v) \over u-v}+\alpha\int_{\CC_+}
dv{\s(v) \over u+v} - \beta\int_{\CC_2}dq{\rho_2(q) \over u-q}\ , }  
\eqn\singeqtwo{0= 2\beta{\int_{\CC_2}\kern-1.5em -}~~dq{\rho_2(q) \over p-q}-{\alpha \over 2}\int_{\CC_+}dv{\s(v) \over p-v}-
{\alpha \over 2}\int_{\CC_+}dv{\s(v) \over p+v}\ , }  where $m$ is an integer enumerating the branch of the logarithm and where
a slash through an integral means a Cauchy principal value prescription. In \singeqone\ and \singeqtwo\ $u,v$ and $p,q$ are the
$u_1$ and $u_2$ roots respectively and we have ignored the $u_3$ roots. By solving \singeqtwo\ for $\r_2$ and plugging the
result into the normalization condition \unity, one can show that the $u_2$ roots distribute themselves on the entire imaginary
axis, \ie $-i\infty<u_2<i\infty$, and that the following relation holds
\eqn\rel{ \int_{-\infty}^{+\infty}dq{\rho_2(iq) \over u-iq}=\int_{\CC_+}dv {\s(v) \over u+v}\ . }
Thus, eliminating $\r_2$ from \singeqone, we are left with the following singular integral equation 
\eqn\matrixmodel{U'(u)=2{\int_a^b\kern-1.4em -}~~dv{\s(v) \over u-v}-n\int_a^bdv{\s(v) \over u+v}\ , }
where $n=-1$. The term on the left hand side of \matrixmodel\ can be thought of as the derivative of a (logarithmic) 
potential $U(u)$ and is for $n=-1$ given by 
\eqn\potone{U'(u)={2 \over \a}\Big({1 \over u}-2\pi m\Big)\ . }
The type of integral equations appearing in \matrixmodel\ are known to arise 
as saddle-point equations in $O(n)$ matrix models, see \eg \refs{\KostovPN,\EynardCN}. By letting $n=2\cos(\pi p/q)$, 
where $p$ and $q$ are positive relatively prime integers,\foot{This parametrization is convenient since $O(n)$ matrix
models are known to exhibit non-trivial critical behaviour only for $-2\leq n \leq 2$. } the solution, expressed
in terms of the resolvent of the eigenvalue density introduced below, is given by a polynomial equation of order
$q$. Hence, for the $O(\pm 1)$ models we expect to find cubic equations.

For a general distribution of Bethe roots on a cut $\CC$, the generating function \gendef\ turns into 
\eqn\therres{t(u)\longrightarrow -{\a \over 2} W(u)\ ,} in the thermodynamic limit, where we have defined the resolvent of
$u_1$ Bethe roots \refs{\EngquistRN,\ArutyunovRG} \eqn\resolvent{W(u)\equiv\int_{\CC} dv{\s(v) \over u-v}\ .} The 
resolvent is analytic throughout the complex plane except across the cut $\CC$. For a {\it symmetric}
distribution of Bethe roots, as the one we are interested in, we can write $t(u)$ as
\eqn\tfcnofw{t(u)=-{\a \over 2}\big(W(u)-W(-u)\big)\ ,} where $W(u)$ is the contribution to the resolvent from the cut $\CC_+$.
Note that for this solution the resolvent of $u_2$ Bethe roots \eqn\resutwo{W_2(p)\equiv\int_{\CC_2} dq{\r_2(q) \over p-q}\ ,}
can be expressed in terms of the resolvent of $u_1$ roots as $W_2(p)=-W(-p)$.
In terms of the resolvent, equation \matrixmodel\ can be written in the compact form
\eqn\reseqn{W(u+i0)+W(u-i0)-W(-u)=U'(u)\ .} We further split the resolvent as $W(u)=W_r(u)+w(u)$, where $w(u)$ solves the
homogeneous part of \reseqn\ and where \eqn\parti{W_r(u)={1 \over 3}(2U'(u)+U'(-u))={2 \over 3\a u} -{4\pi m \over \a}\ ,} is 
a particular solution. Utilizing the method in \refs{\KostovPN,\EynardCN}, we define the even function
\eqn\funr{r(u)\equiv w^2(u)-w(u)w(-u)+w^2(-u)\ ,} which is regular across the cut and can be determined explicitly to be
\eqn\funrtwo{ r(u)={(4\pi m)^2 \over \a^2}+{4 \over 3\a^2u^2}\ ,} by using $W_r(u)$ in \parti, the regularity of the resolvent
\resolvent\ at $u=0$ as well as its asymptotic behaviour. Multiplying $r(u)$ by $w(u)-w(-u)$ we find the equation
\eqn\funs{w^3(u)-r(u)w(u)=-w^3(-u)+r(-u)w(-u)\equiv s(u)\ , } where we have defined the odd function $s(u)$. Since $w(u)$ is
regular for ${\rm Re}(u)<0$ and $w(-u)$ is regular for ${\rm Re}(u)>0$, it follows that $s(u)$ is regular everywhere except
at $u=0$. From \funrtwo\ and the behaviour of $w(u)$ at $u=0$ and $u\rightarrow\infty$, $s(u)$ becomes  
\eqn\sfun{s(u)={16 \over 27\a^3}{1 \over u^3}+8{(2\pi m)^2 \over \a^2}\Big(1-{2 \over 3\a}\Big){1 \over u}\ .}
Using \funr\ and \funs\ one can verify that the difference ${\overline w}\equiv w(u)-w(-u)$ satisfies the cubic equation
\eqn\difference{{\overline w}^3(u)-r(u){\overline w}(u)+s(u)=0\ , } as previously anticipated. Finally, from the relation
\eqn\final{{\overline w}(u)=-{4 \over 3\a u}-{2 \over \a}t(u)\ ,} it is easy to show that $t(u)$ satisfies the cubic
equation \eqn\geneqn{u^2t^3(u)+2ut^2(u)+\big(1-(2\pi m)^2u^2\big)t(u)-\a (2\pi m)^2u=0\ .} This is the main result of this
section. Equation \geneqn\ can easily be solved perturbatively by expanding $t(u)$ around $u=0$. Hence, using \expand\ as an
ansatz the unique odd solution is
\eqn\genofcharge{t(u)=\a(2\pi m)^2u+\a(1-2\a)(2\pi m)^4u^3+\a(1-6\a+7\a^2)(2\pi m)^6u^5+\cdots\ .} At lowest order in
\genofcharge\ we find a charge proportional to the anomalous dimension $\c$ of the operator under consideration, \cf \hamilton. 
In the thermodynamic limit the anomalous dimension as defined in \andimm\ is proportional to the derivative of
the resolvent at $u=0$, so using \resolvent\ and \tfcnofw\ we have 
\eqn\andim{\c\rightarrow{\l \over 8\pi^2J}t'(0)={\l m^2J_2 \over J^2}\ . } Notice that we always have the freedom to
linearly combine the charges, a fact that will be made use of when the comparison with the string theory calculation is made.
 
For the sector in which $J_1<J_2=J_3$ an integral equation similar to \matrixmodel\ is derived, but with
$n=+1$ and a different potential. In \EngquistRN\ it was argued that this sector is the analytic
continuation of the $J_1>J_2=J_3$ sector past an apparent critical point at $J_2=4J_1$, so the
generators resulting from the $O(\pm 1)$ models coincide. From a string theoretic perspective, to which we
turn to in the coming section, this is natural since no critical behavior is seen at
$J_2=4J_1$. 

\newsec{String Theory Calculations}
The goal of this section is to find the generating function of the higher commuting charges from a string theoretic point of
view, following the analysis in \ArutyunovRG. Actually, we will find the charges to all orders in $\l$ (at the semiclassical
level), so the one-loop contribution needs to be extracted in order to compare with the gauge theory
result. We start by reviewing some of the features of the three-spin string solution with $R$-charge
assignments $(J_1,J_2,J_2)$ on $S^5$ first considered in \FrolovQC\ and further developed in \FrolovTU. From the
B\"acklund transformations
\refs{\PohlmeyerNB,\OgielskiHV,\ArutyunovRG}\ for the $O(6)$ sigma model we find a one-parameter family of 
string solutions given as a function of a certain spectral parameter by treating the three-spin solution as the trial
solution. We find that the redefinition of charges on the string side is exactly the same as in \ArutyunovRG. However, using 
the freedom to linearly combine the charges, we find it more convenient to work with an ``improved'' spectral parameter
introduced below at the cost of slightly complicating the equations but making the comparison with the gauge theory
calculations more direct.

\subsec{The Frolov-Tseytlin String Solution}
We begin with summarizing the part of the $(J_1,J_2,J_2)$ solution relevant for the present analysis and refer
the reader to \refs{\FrolovQC,\FrolovTU}\ for details. We consider a closed string at the center of
AdS$_5$ ($\r=0$ in global coordinates) rotating in three orthogonal planes on the five-sphere. The
AdS$_5$ time coordinate is denoted by $t=X_0(\t,\s)$ and we work with the complexified five-sphere
embedding coordinates $Z_I(\t,\s)\equiv X_{2I-1}(\t,\s)+iX_{2I}(\t,\s)$, $I=1,2,3$, where $(\t,\s)$ parametrize
the string world-sheet. Using conformal gauge, the equations of motion and the conformal gauge constraints for
strings on the five-sphere are given in world-sheet light-cone coordinates $\xi=(\t+\s)/2$, $\eta=(\t-\s)/2$ by
\eqn\eom{\eqalign{&Z_{\xi\eta}+{\rm Re}(Z_\xi\cdot{\bar Z}_\eta)Z=0\ , 
\qquad Z\cdot{\bar Z}=1\ , \cr &Z_\xi \cdot {\bar Z}_\xi=t_\xi^2\ , \qquad 
Z_\eta \cdot {\bar Z}_\eta=t_\eta^2\ ,   } }
where $Z_\xi\equiv\partial_\xi Z$ and $Z_\eta\equiv\partial_\eta Z$. The closed string periodicity condition
is also required. The $(J_1,J_2,J_2)$ solution is given by:
\eqn\trial{\eqalign{t&=\k\t\ , \cr Z_I(\t,\s)&
=u_I(\s)\exp(iw_I\t)\ , \qquad I=1,2,3\ , } }
where two of the angular velocities are equal, say $w_2=w_3=w$ and $w_1=\nu$, and where $u_I(\s)$ parametrizes the unit
$2$-sphere \eqn\trialu{\eqalign{u_1(\s)&=\cos\c_0 \ , \cr
 u_2(\s)&=\sin\c_0\cos m\s \ , \cr
 u_3(\s)&=\sin\c_0\sin m\s\ . } }
Here $\k$, $\c_0$ and $m$ are constants. From the equations
of motion and the conformal gauge constraints in \eom\ a set of relations is obtained
between the parameters given in the solution \trial\ and \trialu\
\eqn\confeom{w^2=\nu^2+m^2\ , \qquad \k^2=\nu^2+2m^2\sin^2\c_0\ ,
\qquad m\in{\hbox{ Z\kern-.8em Z}}\ .}
In the following we will treat $\k$ and $\nu$ as independent 
parameters. Their expansions in powers of the effective string coupling $\l'=\l/J^2$
are given in Appendix A. The space-time energy of the string is expressed in terms of $\k$ as 
\eqn\ener{E=\sqrt{\l}\k=J+{\l m^2J_2 \over J^2}+\cdots\ ,} 
which via the AdS/CFT correspondence is identified with the scaling dimension
of the corresponding operator \operator\ in the gauge theory. To lowest order we recognize the canonical
dimension counting the number of constituent fields and to first order in $\l/J^2$ the
one-loop anomalous dimension, \cf \andim. 

\subsec{The B\"acklund Transformations} 
It is well-known that $O(n)$ sigma models are classically integrable, see \eg \refs{\PohlmeyerNB{,}\OgielskiHV}, 
which in particular implies the existence of an infinite number of conserved charges. One way of constructing 
{\it local} conservation laws is to use parametric B\"acklund transformations. Given a ``trial'' solution
$Z(\t,\s)$ to the classical equations of motion in \eom\ a one-parameter family of solutions $Z^{(\c)}(\t,\s)$  
with $\c\in{\rm I\kern-.18em R}$ also solving \eom\ can be obtained by requiring a certain set of B\"acklund 
equations be satisfied. This ``dressed'' solution can then be used to construct the generator of conserved charges.  
     
At the cost of making the equations look a little bit more complicated one can make the correspondence between the 
gauge theory and the string theory calculations more explicit by expanding the dressed solution not in
the spectral parameter $\c$ usually considered, see \eg \refs{\PohlmeyerNB,\OgielskiHV,\ArutyunovRG},
but in an ``improved'' spectral parameter \eqn\imprspect{\m={\c \over 1+\c^2}\ , \qquad\qquad
|\m|\leq1/2\ ,} also considered in \ArutyunovRG\foot{In \ArutyunovRG\ the relation \imprspect\ was presented as
$1-4\mu^2=(1-\c^2)^2/(1+\c^2)^2$.}. Inverting this quadratic equation in $\c$ yields
\eqn\inv{\c={1 \over 2\m}(1-\sq)\ ,}
where the requirement that $\c=0$ should correspond to $\m=0$ singles out one solution and consequently excludes the
region $|\c|>1$. The B\"acklund equations determining the dressed string solution
$Z_I^{(\m)}$ then take the form \OgielskiHV
\eqn\backl{\eqalign{ (1-\sq)(Z_I^{(\m)}+Z_I)_\xi&=
+{\rm Re}(Z^{(\m)}\cdot {\bar Z}_\xi)(Z_I^{(\m)}-Z_I)\ ,\cr
{4\m^2 \over 1-\sq}(Z_I^{(\m)}-Z_I)_\eta&=-{\rm Re}(Z^{(\m)}\cdot {\bar Z}_\eta)
(Z_I^{(\m)}+Z_I)\ ,} }
together with the normalization and ``initial'' conditions 
\eqn\incond{ Z^{(\m)}\cdot{\bar Z}^{(\m)}=1\ , \quad  Z_I^{(\m)}(\t,\s){\big |}_{\m=0}=Z_I(\t,\s)\ , }
\eqn\normcond{{\rm Re}(Z^{(\m)}\cdot{\bar Z})=\sq\ ,} 
where $Z_I(\t,\s)$ is any trial solution solving the classical equations of motion \eom. 

Knowing the dressed solution $Z^{(\m)}$ the generator of local commuting charges $\cE(\m)=\sum_m\cE_m\m^m$ can be
found from\foot{$\cE(\m)$ is related to $\cE(\c)$ in \ArutyunovRG\ by the redefinition $\cE(\m)=\cE(\c)/\c^2$.} 
\eqn\energysg{\cE(\m)={1 \over 4\pi\m(1-\sq)} \int d\s {\rm Re}\Big(
2\m^2Z^{(\m)}\cdot{\bar Z}_{\xi}+(1-2\m^2-\sq)Z^{(\m)}\cdot{\bar Z}_{\eta}\Big)\ , }
whose conservation law can be derived from \backl--\normcond.

\subsec{Finding the Generating Function}
Next, we determine the generating function $\cE(\m)$ of higher charges evaluated on the classical solution \trial\
to an arbitrary order in the spectral parameter $\m$ and to all orders in $\l=R^4/\a'^2$ in the strict
$J\rightarrow\infty$ limit in which quantum corrections are suppressed. To achieve this we 
make a certain ansatz of the dressed solution, first considered in \ArutyunovRG, plug it into
\energysg\ and then use the B\"acklund transformations \backl\ to show that $\cE(\m)$ satisfies a sixth order
equation (recall that $t(u)$ satisfy the cubic equation \geneqn). In order to match this
generator with the generator $t(u)$ found in the gauge theory \genofcharge\ we extract the one-loop
contribution by a certain rescaling of $\cE(\m)$ and by letting $\cJ=J/\sqrt{\l}$ approach infinity, a
procedure very reminiscent of that in \ArutyunovRG. For details we refer the reader to Appendix B.

Inspired by \ArutyunovRG, we make the following ansatz for the dressed solution 
\eqn\dress{Z_I^{(\m)}(\t,\s)\equiv X_{2I-1}^{(\m)}+iX_{2I}^{(\m)}
=r_I(\s,\m)\exp\big(i\a_I(\m)\big)\exp(iw_I\t)\ , } where
\eqn\uandr{\eqalign{r_1(\s,\m)&=\cos\c_0 \ , \cr
 r_2(\s,\m)&=\sin\c_0\cos\big(m\s+\th(\m)\big) \ , \cr
 r_3(\s,\m)&=\sin\c_0\sin\big(m\s+\th(\m)\big)\ . } }
Here the shift function $\th(\m)$ and the phases $\a_I(\m)$ are functions of only the spectral
parameter $\m$ and vanish for $\m=0$, \ie $Z^{(\m)}$ reduces to $Z$ for $\m=0$. The one-parameter family of solutions in 
\dress\ solves the equations of motion (or equivalently the equations of motion of the NR system  \ArutyunovZA\ with constant
$\Lambda$ and $m_2=-m_3=m$, $m_1=0$) 
and the conformal gauge constraints in \eom\ for arbitrary $\m$ provided that the relations in \confeom\ hold.

Substituting the ansatz \dress\ into \energysg\ and performing some elementary integrals the 
generating function becomes
\eqn\energysgg{\cE(\m)={1 \over 2\m(1-\sq)}{\rm Re}\Big(
2\m^2Z^{(\m)}\cdot{\bar Z}_{\xi}+(1-2\m^2-\sq)Z^{(\m)}\cdot{\bar Z}_{\eta}
\Big){\Big |}_{\s=0}\ . }
By further plugging in the ansatz \dress\ into the B\"acklund equations \backl\ yields an
overdetermined system of equations out of which four are independent. This determines the functions
$\a_I(\m)$, $I=1,2,3$ and $\th(\m)$ completely. Solving \backl\ for
${\rm Re}(Z^{(\m)}\cdot{\bar Z}_{\xi})$ and ${\rm Re}(Z^{(\m)}\cdot{\bar Z}_{\eta})$, substituting the
result into \energysgg, the generator
takes the very simple form \eqn\energysggg{\cE(\m)=\m {2\nu \over \sin\a_1}\ . }
Now from \backl\ and \normcond\ $\sin\a_1$ can be shown to satisfy 
a sixth order equation which consequently determines an equation for the generating function 
\eqn\gensixth{4\nu^2\mu^4\vt^2+\cE^2\mu^2\big(4m^4(1-4\m^2)+\vt(4\nu^2-\vt)\big)+
\cE^4(\k^2-8m^2\mu^2)-\cE^6=0\ ,}
where for notational simplicity we have defined
\eqn\func{\vt\equiv2m^2+\nu^2-\k^2=2m^2\cos^2\c_0\ .}
Let us consider some limiting cases. When $J_1=0$ the string reaches its maximal size ($\c_0=\pi/2$) implying that
the function $\vt$ vanishes and \gensixth\ simplifies to a quartic equation. This corresponds to a string 
carrying the representation $[J_2,0,J_2]$\foot{This (unstable) circular two-spin solution \FrolovQC\ is a limit of the more
general two-spin solution of the NR system with the shape of a {\it bent} circle considered in 
\refs{\ArutyunovUJ,\ArutyunovRG}.} and was analyzed on the gauge theory side in \BeisertXU.
Considering instead the limit $J_2=J_3=0$ we recover the BMN state \BerensteinJQ\ in the representation 
$[0,J_1,0]$ in which the string is point-like. For this solution $\k=\n$ and $m=0$, so also here $\vt$ vanishes. 
Hence, \gensixth\ shows that the generating function $\cE(\m)\rightarrow\k$ and no higher charges are present. 

Using the expansions of $\k$ and $\nu$ in powers of $\l/J^2$ given in Appendix A, the equation 
\gensixth\ can be solved perturbatively in (even) powers of $\mu$
\eqn\eqalmim{\cE(\m)=\sum_{m=0}^{\infty}\cE_{2m}\mu^{2m}\ .}
The resulting expressions for the for the first few charges
$\cE_m=\cE_m(\k,\n)$ evaluated on our three-spin solution are given in Appendix B. Expanding in $\l/J^2$ to first
order we find \eqn\result{\eqalign{\sqrt{\l}\cE_0&=J+{\l m^2J_2 \over J^2} + \cdots\ , \cr 
\sqrt{\l}\cE_2&=-{2^3\l m^2J_2 \over J^2}+\cdots\ ,  \cr 
\sqrt{\l}\cE_4&=+{2^5\l^2 m^4J_2 \over J^5}(J-4J_2)+\cdots\ , \cr 
\sqrt{\l}\cE_6&=-{2^7\l^3 m^6J_2 \over J^8}(J^2-12J_2J+28J_2^2)+\cdots\ , \cr
\vdots &}  }
These charges satisfy ``BMN scaling'' \refs{\BerensteinJQ,\ArutyunovRG}\ where the {\it m}-th charge scales as $\cJ^{1-m}$. We
observe that the charges $t_{2m+1}$, $m=0,1,2,\ldots$, found in \genofcharge\ in the gauge theory appear as the {\it leading 
order} terms in $\cE_{2m+2}$. The one-loop generating function $\cQ(u)$ is then easily extracted from $\cE(\m)$ by 
the following limiting procedure\foot{This closely resembles the limiting procedure in \ArutyunovRG\ for the 
two-spin case, the main difference being that $\cQ(u)$ here is rescaled by a factor of $u$. Also, 
$\cE(\m)$ given here differs from ${\widetilde\cE}(\m)$ in \ArutyunovRG\ by a factor of $\m^2$.}
\eqn\genchfrst{u\cQ(u)\equiv\lim_{\cJ\rightarrow\infty}\Big({\cE(\mu) \over \cJ}
-1\Big)\ , } where we have identified the spectral parameter of the gauge theory generating function 
\eqn\identbethe{u^2=-{\mu^2 \over \pi^2\cJ^2}\ . }
In taking this limit we have lost information about the zeroth order charge, proportional to the energy. 
To provide ourselves with a ``proof'' of the one-loop correspondence at a functional level we take the limit
\genchfrst\ directly in \gensixth, and find that the left hand side of the sixth order equation factorizes into two 
cubic factors: \eqn\factorize{\eqalign{&\Big[\cQ(u\cQ+1)^2-(2\pi m)^2u(u\cQ+\a)\Big] \cr 
&\times \Big[u^3\cQ^3+4u^2\cQ^2+u\cQ\big(5-(2\pi m)^2u^2\big)+(2\pi m)^2u^2(\a-2)+2\Big]=0\ ,} } 
where we have used the ``filling fraction'' $\a=2\cJ_2/\cJ$ previously defined. The first square
bracket we recognize as the left hand side of the cubic equation \geneqn, since $\cQ(u)$ can be identified with
$t(u)$! The second factor in \factorize\ is related to the first by the discrete transformation 
\eqn\discrete{\cQ(u)\longrightarrow-{2 \over u} -\cQ(u)\ ,} and appears to yield no new information.

We conclude this section by presenting an explicit expression for the one-loop generating function 
appropiate for computing the charges perturbatively. From \energysg\ and \genchfrst\ an expansion around $\m=0$ yields
\eqn\final{u\cQ(u)=\lim_{\cJ\rightarrow\infty}{1 \over 4\pi\m\cJ}\int d\s{\rm Re}\Big((Z^{(\m)}\cdot{\bar Z}_{\xi})
-2\m\cJ+\big((Z^{(\m)}\cdot({\bar Z}_{\eta}-{\bar Z}_{\xi})\big)\cP(\m)\Big)\ ,  }
where $\cP(\m)$ is the polynomial
\eqn\pol{\cP(\m)=\m^2+\m^4+2\m^6+5\m^8+14\m^{10}+42\mu^{12}+\cdots\ .} The expansions of 
${\rm Re}(Z^{(\m)}\cdot{\bar Z}_{\xi})$ and ${\rm Re}(Z^{(\m)}\cdot{\bar Z}_{\eta})$ in powers of $\m$ and $1/\cJ$ 
are given in Appendix B for the $(J_1,J_2,J_2)$ solution. 

\newsec{Conclusions}
The main objective of this study was to gain further clues about the relation between the integrability on the two
sides of the planar AdS/CFT duality by considering a specific example, using the method initiated in \ArutyunovRG. Recall that
integrability appears very differently on
the two sides: on the gauge theory side we consider a quantum integrable spin chain, whereas on the string theory side 
we consider a classical string sigma model. By dressing a specific three-spin solution in type IIB string theory we found a
one-parameter family of string solutions
$Z^{(\m)}$ and further derived a function $\cE(\m)$ generating an infinite tower of charges in involution. To
first order in the expansion parameter $1/\cJ$ these were successfully matched onto the gauge theory result found
in \EngquistRN. 

We have gained confidence that the limit in \genchfrst\ is the way to extract the one-loop contribution from the string theory
generator of higher charges, \cf \ArutyunovRG\foot{Note the similarity of the result in the present paper with the ones in 
\ArutyunovRG. In particular, if the generator is expanded in powers of $\c$ instead of $\m=\c/(1+\c^2)$, 
${\widehat \cE}(\c)=\sum_{m=2}^{\infty}{\widehat \cE}_m\c^m$, the {\it same} order by order redefinitions of the charges as in
\ArutyunovRG\ are required to produce the improved charges.}.
This strongly suggests that there should be a more general prescription relating the generating functions on the two 
sides of the duality in the semiclassical (thermodynamic) limit, not specific to any particular solution.  Eq.~\final\ should
be regarded as an explicit mapping relating the higher charges on the two sides of the duality, given that a string 
solution is known. Furthermore, a map relating the B\"acklund equations and the Bethe equations in the semiclassical limit is
suggestive.

At the time of this writing there is evidence that perturbative integrability is valid in the planar limit to
two loops in $\cN=4$ SYM, see for instance \refs{\BeisertTQ,\SerbanJF}. Further conjectures of higher-loop
integrability on the gauge theory side have been put forward in \eg \refs{\BeisertJB,\BeisertYS,\BeisertEA}, assuming 
perturbative BMN scaling. On the string theory side we know that classical integrability of the $SO(4,2)\times SO(6)$ sigma
model holds to all orders. With this in mind, it would be of interest to generalize the results on the gauge theory side to 
higher orders. To two loops we expect that operators containing only scalars start to mix with operators containing fermions
and field strengths. One way to proceed is to calculate the generating function of conserved higher charges using an analog of 
the integrable Inozemtsev long range spin chain \SerbanJF.   

\bigbreak\bigskip\bigskip\centerline{{\bf Acknowledgements}}\nobreak 
I am grateful to J.~Minahan and K.~Zarembo for discussions and comments on the manuscript and to L.~Freyhult and 
M.~Smedb\"ack for discussions. 

\appendix{A}{Expansions of $\k$ and $\nu$}
For the $(J_1,J_2,J_2)$ solution three components of the $SO(6)$ angular momentum tensor $J_{AB}$, $A,B=1,\ldots,6$ are 
non-zero, corresponding to rotations on the five-sphere in three orthogonal planes:
\eqn\ang{J_1=J_{12}\ , \qquad J_2=J_{34}\ , \qquad J_3=J_{56}\ .} We recall \FrolovTU\ that the only restriction on the
$R$-charges in this solution comes from the stability requirement $J_2+J_3\leq(4m-1)J_1/(2m-1)^2$, where $m$ is the winding 
number. Evaluating these components using the solution \trial\ and the definitions
\eqn\angular{J_I\equiv{i\sqrt{\l} \over 4\pi}\int d\s \big(Z_I\partial_{\tau}{\bar Z}_I-{\bar Z}_I\partial_{\tau}Z_I\big)\ , 
\qquad ({\rm no~sum})\ } one easily infers the following relations between the space-time energy of the
string $E=\sqrt{\l}\k$ and the ``charge'' $\cV=\sqrt{\l}\nu$ \refs{\FrolovQC,\FrolovTU}\
\eqn\enandnuone{E^2=\cV^2\Big(1+{2\l m^2 \over \cV^2}\big(1-{J_1 \over \cV}\big)\Big)\ ,}
\eqn\enandnutwo{(\cV-J_1)^2\big(1+{\l m^2 \over \cV^2}\big)=4J_2^2\ ,} where $E$ has been eliminated in the last equation using
\enandnuone.
Solving these equations as an expansion in powers of  $\l'=R^4/\a'^2J^2$ and the rescaled charges $\cJ=J/\sqrt{\l}$ and
$\cJ_2=J_2/\sqrt{\l}$ yield 
\eqn\kexp{\eqalign{\k&=\cJ+\l' m^2\cJ_2-{\l'^2m^4\cJ_2 \over 4}+
{\l'^3m^6\cJ_2 \over 8} \Big(1-{4\cJ_2 \over \cJ}+{8\cJ_2^2 \over \cJ^2}\Big) + \cdots\ , \cr
\n&=\cJ-\l' m^2\cJ_2+{\l'^2m^4\cJ_2 \over 4}
 \Big(3-{8\cJ_2 \over \cJ} \Big)+{\l'^3m^6 \over 8} \Big({36\cJ_2 \over \cJ}-5-{56\cJ_2^2 \over \cJ^2}\Big)
+\cdots\ . }  }

\appendix{B}{Details of the B\"acklund Solution and the Generating Function}
Let us collect the main ingredients needed to derive equations \energysggg\ and \gensixth, valid for the $(J_1,J_2,J_2)$ 
solution. Putting $m\s=0,\pi/2$ in \normcond\ we find the conditions that $\a_2=\a_3$ and that 
\eqn\normcon{\cos^2\c_0\cos\a_1+\sin^2\c_0\cos\th\cos\a_2=\sq\ .} 
The information that we need from the six components of the complex B\"acklund equations in \backl\ is most
easily extracted by taking the real parts and putting $\s=0$. After some algebra we get the following relations
between the $\m$-dependent functions $\th(\m)$ and $\a_I(\m)$
\eqn\thetafcns{\eqalign{&\cos\th={\nu \over \sqrt{m^2\sin^2\a_1+\nu^2}}\ ,
\qquad ~~\sin\th={m\sin\a_1 \over \sqrt{m^2\sin^2\a_1+\nu^2} } \ , \cr
& w^2\cos^2\th=\nu^2+m^2\cos^2\a_2\ , \qquad w\sin\th=m\sin\a_2\ . } }
The imaginary parts can be shown to give rise to no additional information. To derive \energysggg\ we need the integrals
\eqn\rela{\eqalign{\int_0^{2\pi} d\s{\rm Re}(Z^{(\m)}\cdot{\bar Z}_{\xi})=2\pi{\rm Re}(Z^{(\m)}\cdot{\bar Z}_{\xi})
{\big |}_{\s=0}\ , \cr \int_0^{2\pi} d\s{\rm Re}(Z^{(\m)}\cdot{\bar Z}_{\eta})=2\pi{\rm Re}(Z^{(\m)}\cdot{\bar Z}
_{\eta}){\big |}_{\s=0}\ , } } as well as the expressions
\eqn\rez{\eqalign{
& {\rm Re}(Z^{(\m)}\cdot{\bar Z}_{\xi}){\big |}_{\s=0}=\nu\cos^2\c_0\sin\a_1+\sin^2\c_0(m\cos\a_2\sin\th
+w\sin\a_2\cos\th)\ , \cr
& {\rm Re}(Z^{(\m)}\cdot{\bar Z}_{\eta}){\big |}_{\s=0}=\nu\cos^2\c_0\sin\a_1+\sin^2\c_0(-m\cos\a_2\sin\th
+w\sin\a_2\cos\th)\ , } }
which have the following expansions in powers of $1/\cJ$ and $\m$:
\eqn\ted{\eqalign{ & {\rm Re}(Z^{(\m)}\cdot{\bar Z}_{\xi}){\big |}_{\s=0}=\Big(2\cJ+{2m^2\cJ_2 \over \cJ^2}+\cdots
\Big)\m \cr & \qquad\qquad\qquad\qquad\quad+ \Big(-{8m^2\cJ_2 \over \cJ^2}+{12m^4\cJ_2 \over \cJ^4}
\Big({4\cJ_2 \over \cJ}-1\Big)+\cdots\Big)\m^3+\cdots\ , \cr & {\rm Re}(Z^{(\m)}\cdot{\bar Z}_{\eta}){\big |}_{\s=0}=
\Big(2\cJ-{6m^2\cJ_2 \over \cJ^2}+\cdots\Big)\m \cr & \qquad\qquad\qquad\qquad\quad+ \Big({8m^2\cJ_2 \over \cJ^2}
+{4m^4\cJ_2 \over \cJ^4}\Big(3-{20\cJ_2 \over \cJ}\Big)+\cdots\Big)\m^3+\cdots\ . } } 
Eliminating $\cos\a_2$ and 
$\cos\th$ from \normcon\ gives an equation for $\a_1$. In particular, by squaring \normcon, one can eliminate the
$\cos\a_1$ term and express all $\a_1$-dependence in powers of $\sin\a_1$:
\eqn\sineq{\eqalign{& m^4\cos^4\c_0\sin^6\a_1+m^2\sin^4\a_1\Big(m^2(1-4\m^2)+2\nu^2\cos^2\c_0-m^2\cos^4\c_0\Big)
\cr & +\nu^2\sin^2\a_1\Big(2m^2(1-4\mu^2)+\nu^2-2m^2\cos^2\c_0\Big)-4\mu^2\nu^4=0\ . } }
Finally, we get \gensixth\ by expressing $\cos\c_0$ in terms of $\k$ and $\nu$ using \confeom\ and expressing $\sin\a_1$ in 
terms of $\cE(\m)$ using \energysggg. Note that to first order
in $\m$ \sineq\ gives that $\sin\a_1=2\m\nu/\k+\cO(\m^2)$ which from \energysggg\ implies that $\lim_{\m\rightarrow 0}\cE(\m)=\k$
as it should. 

Let us also summarize the first few charges evaluated on our three-spin solution, as functions of $\k$ and $\nu$. 
These can be determined either by solving the B\"acklund equations \backl\ perturbatively as in Section 3.3 of
\ArutyunovRG, a procedure with a rapidly increasing degree of difficulty, or by solving \gensixth\ perturbatively. 
Either way the lowest-lying charges become
\eqn\resultexact{\eqalign{\cE_0=&\k\ , \cr 
\cE_2= &{(\nu^2-\k^2) \over 2\k^3} (\k^2-3\n^2+4m^2)\ ,\cr
\cE_4= &{(\nu^2-\k^2) \over 8\k^7}\Big((\k^2-\nu^2)(5\k^4+14\k^2\nu^2+45\nu^4) \cr 
&\qquad\quad\quad~~+m^2(8\k^4+48\k^2\nu^2-120\nu^4)+m^4(16\k^2-80\nu^2)\Big)\ , \cr
\cE_6= & {(\nu^2-\k^2) \over 16\k^{11}}\Big(7(\nu-\k)^2(\nu+\k)^2(\k^2+3\nu^2)(3\k^4+2\nu^2\k^2+27\nu^4)
\cr &\qquad\quad\quad~~+4m^2(\nu^2-\k^2)(567\nu^6-133\k^2\nu^4-43\k^4\nu^2-7\k^6) \cr 
&\qquad\quad\quad~~+m^4(240\k^4\nu^2-2800\k^2\nu^4+3024\nu^6+48\k^6) \cr
&\qquad\quad\quad~~+m^6(64\k^4-896\k^2\nu^2+1344\nu^4) \Big)\ ,   } }
where the factorization ($\nu^2-\k^2$) for the charges $\cE_m$, $m\ge 2$ ensures that there are no higher charges for the
BPS protected point-like string state \BerensteinJQ, for which $\nu=\k$.

\listrefs

\bye